\documentclass[english,french,epj]{svjour}
\usepackage[latin9]{inputenc}
\usepackage{amsmath}
\usepackage{graphicx}
\usepackage{amssymb}
\usepackage{esint}

\usepackage{ulem}
\institute{Laboratoire PhLAM, CNRS UMR 8523, Bât. P5 - Université Lille1, 59655 Villeneuve d'Ascq cedex, FRANCE}
\abstract{An exhaustive kinetic model for the atoms in a 1D Magneto-Optical Trap is derived, without any approximations. It is shown that the atomic density is described by a Vlasov-Fokker-Planck equation, coupled with two simple differential equations describing the trap beam propagation. The analogy of such a system with plasmas is discussed. This set of equations is then simplified through some approximations, and it is shown that corrective terms have to be added to the models usually used in this context.
}

\usepackage{babel}
\addto\extrasfrench{\providecommand{\fg}{\ifdim\lastskip>\z@\unskip\fi~\frqq}}

\begin{document}

\title{Phase-space description of the magneto-optical trap}

\author{Rudy Romain, Daniel Hennequin and Philippe Verkerk}

\maketitle

\section{Introduction}

The magneto-optical trap (MOT) is the primary tool to cool atoms.
The development of this technique led to spectacular breakthroughs
in experimental quantum physics: MOTs are the first step in the realization
of optical lattices \cite{lattices}, cold molecules \cite{molecules},
or Bose-Einstein condensates \cite{BEC}. But the MOT is also an interesting
object \textit{per se}. It produces a cloud of cold atoms, the physics
of which is complex. In particular, spatio-temporal instabilities
of this cloud are commonly observed \cite{wilkowski2000,labeyrie2006}.
Several models with different approaches have been proposed to describe
these dynamics, and to identify the mechanisms leading to instabilities.
Unfortunately, none of these models gives a satisfying description
of the observed dynamics. In \cite{wilkowski2000}, a very simple
model allowed to describe experimentally observed instabilities. This
model, improved in \cite{hennequin2004}, predicted another type of
instabilities, which were effectively observed in \cite{distefano2003,distefano2004}.
However, the agreement between the model and the experiments was only
qualitative. Moreover, it concerned the particular case where the
counterpropagating trap beams are obtained by retro-reflection, i.e.
a global asymmetry is introduced in the trap.

Recently, it has been proposed to describe the symmetric MOT, with
all trap beams which are independent, as a weakly damped plasma \cite{labeyrie2006,pohl2006}.
Indeed, the cloud of cold atoms in a MOT is a confined dilute object
with long-range interactions, as in plasmas \cite{walker1990,pruvost2000}.
This model predicted the existence of instabilities above a relatively
high threshold, so that instabilities should exist only in large MOTs.
This seems in contradiction with the observations related in \cite{hennequin2004}.
Moreover, no direct comparison with the experimental temporal regimes
allowed validating this model. More recently, a more complete description
was derived using the methods of waves and oscillations in plasmas,
leading to interesting predictions \cite{mendonca2008}. But as in
the previous case, this study is based upon intermediate well-established
results, valid only in specific cases (e.g. a low beam intensity or
a negligible viscosity). These conditions do not correspond in general
to the experimental situations, and indeed, these results were not
compared to experimental results.

It appears from these numerous works that a reference model for MOT
atom clouds lacks. Such a model should be as general as possible,
and should at least describe the usual experimental situations. In
particular, such approximations as the low saturation limit should
not be done \textit{a priori}. Another interest of such a model is
that it could help in determining precisely the analogies between
MOT atom clouds and other systems, such as plasmas.

The present work is a first step towards such a model. Its aim is
to build a model, with the least possible hypotheses and approximations,
of a 1D symmetric MOT. The resulting set of equations describes as
precisely as possible the dynamics of atoms inside a 1D MOT, and constitutes
a basis model. If simplifications are necessary, approximations should
be applied on this model, \textit{a posteriori.}

The paper is organized as follows: section 2 defines the bases of
the model, while in section 3, we show that the dynamics of the MOT
phase space density can be described by a Vlasov-Fokker-Planck equation
with two relaxation processes. In section 4, we derive the different
terms of this equation of evolution as a function of the usual experimental
parameters. In section 5, we establish the equations of propagation
of the trapping beams. We obtain a set of coupled equations fully
describing our system, in the general case. This set of equation can
be solved numerically to obtain solutions of the general system. However,
to have a better understanding of the MOT, it is interesting to go
further in the analytical approach, simplifying \textit{a posteriori}
the model, as discussed above. This is done in section 6, where we
examine an approximation for the atomic response.

\section{Definition of the model}

\begin{figure}
\includegraphics[width=7.5cm]{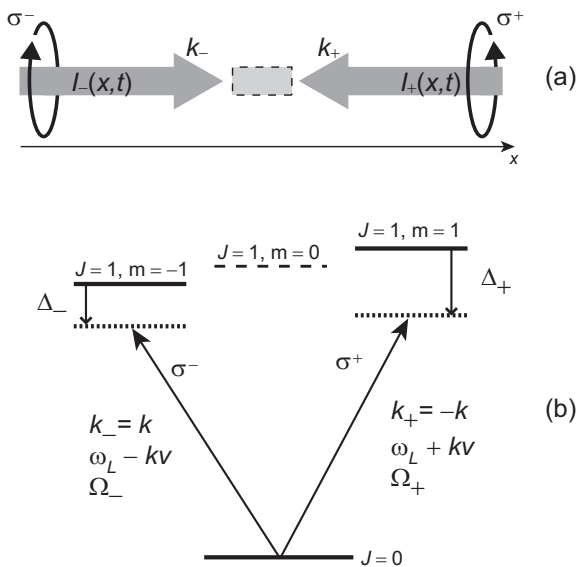}\caption{\label{fig:Sketch}Sketch of the 1D system considered here. a) the
two contrapropagating beams have opposite circular polarizations.
b) The laser beams interact with {}``three-level'' atoms on a $J=0\rightarrow J'=1$
transition, the degeneracy of which is lifted by an inhomogeneous
magnetic field. $\Omega_{\pm}$ are the Rabi frequencies associated
with the beams; $\Delta_{\pm}$ are the effective detunings (see section
4.1).}

\end{figure}

As discussed above, we consider here a 1D configuration. Two counterpropagating
laser beams with opposite circular polarizations interact with the
atoms, as sketched on Figure \ref{fig:Sketch}a. The beam with the
$\sigma^{-}$ polarization comes from the negative abscissa, and is
denoted by the minus sign (intensity $I_{-}(x,t)$, wave vector $k_{-}$).
In the same way, the beam with the $\sigma^{+}$ polarization comes
from the positive abscissa (intensity $I_{+}(x,t)$, wave vector $k_{+}$).
Forces originate from the exchange of momentum between the atoms and
the electromagnetic field. We consider here that the atoms are the
simplest ones for which the magneto-optical trapping is possible.
The laser frequency $\omega_{L}$ is tuned in the vicinity of a $J=0\rightarrow J'=1$
transition with a frequency $\omega_{0}$ (Fig. \ref{fig:Sketch}b).

In the 2D phase space, a point has the coordinates $\left(x,p\right)$,
where $x$ is the position and $p$ is the momentum. To describe the
cloud of cold atoms, we introduce the phase space density $\rho\left(x,p,t\right)$.
Formally, the complete atomic system is described by a phase space
density including both the hot and the cold atoms. However, laser
cooling acts only in a limited region of space, typically the intersection
of the laser beams, and for moderate velocities. As a consequence,
the surrounding hot atoms are considered as a large reservoir which
remain in a thermal equilibrium at room temperature. In the following,
we neglect the action of the cold atoms on the hot background. These
approximations lead to a huge simplification in the description of
the collisions. Thus, to describe the cold atom dynamics, we have
just to derive the equation of evolution of $\rho\left(x,p,t\right)$
from the basic principles of atomic physics. This approach allows
us to go beyond the Wieman description of multiple scattering in terms
of absorption cross sections \cite{walker1990,sesko1991}. $\rho\left(x,p,t\right)$
influences the propagation of the trapping beams, and reciprocally
the modification in the beam intensities changes the evolution of
the phase space density. Thus, we expect to obtain a system of coupled
nonlinear differential equations. As our aim is to build a theoretical
frame in which most of the experimental situations can be explored,
we limit as much as possible the initial approximations. In particular,
we do not restrict our study to the low saturation limit, as in \cite{mendonca2008}.

To derive the system of coupled equations, we proceed in two steps.
We first assume that we know the intensities of the beams everywhere
in the sample, and we derive the equation of evolution of the density
in phase space (section 3 et 4). Then, we write the equations of propagation
of the beams assuming that we know the atomic density in phase space
(section 5).

\section{Evolution of the phase space density}

To derive the equation of evolution of the phase space density $\rho\left(x,p,t\right)$,
we consider an elementary cell centered in $\left(x_{0},p_{0}\right)$,
with dimensions $\delta x$ and $\delta p$. The number of atoms contained
in this cell is $N(t)=\rho\left(x_{0},p_{0},t\right)\delta x\,\delta p$,
where we assume that $\delta x$ and $\delta p$ have been chosen
small enough to neglect higher order corrections. The variations of
$N$ between $t$ and $t+\delta t$ are governed by three distinct
phenomena: (i) an atom is kicked in or out by a collision with the
hot background gas or (ii) an atom crosses the border of the cell
because of the evolution of either its position or (iii) its momentum.
The simultaneous change in position and in momentum would lead to
second order terms in $\delta t$, which are neglected.

The collisional processes lead to two terms, one for losses and one
for gains : \begin{equation}
\delta N_{coll}=-\frac{N\,\delta t}{\tau}+\lambda\,\delta x\,\delta p\,\delta t\label{eq:dNcoll}\end{equation}
where $\tau$ is the mean time interval between two collisions with
the hot atoms. The source term $\lambda$ is due to the collisions
between two hot atoms that tend to restore the Maxwell-Boltzmann distribution
for the phase space density (uniform in $x$, gaussian in $p$). A
contribution from the collisions between cold atoms could also be
considered but, in the conditions where instabilities are observed
in a MOT, this last contribution can be neglected. 

\begin{figure}
\includegraphics[width=8cm]{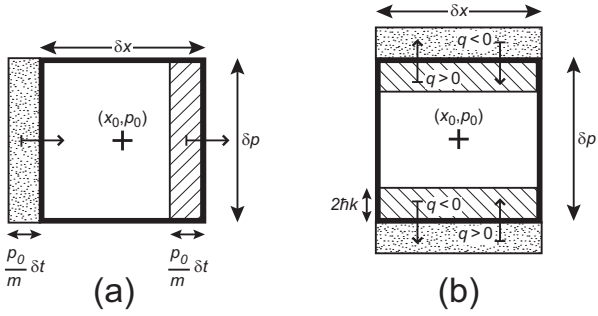}\caption{\label{fig:Cell}Modification of the number of atoms contained in
the elementary cell due to the evolution of (a) the position and (b)
the momentum. The dashed regions contribute to losses, while dotted
regions contribute to gains. In (a), all the atoms in the indicated
regions will cross a boundary (we assume that $p_{0}>0$), and the
areas are proportional to $\delta t$. In (b), a small fraction of
the atoms, proportional to $\delta t$, will receive a momentum kick.
We have to consider separately the momentum kicks with $q>0$ and
those with $q<0$.}

\end{figure}

The second mechanism is the variation of $N$ due to the velocity
of the atoms. As depicted in Fig. \ref{fig:Cell}a, the atoms that
cross the borders in $x_{0}\pm\delta x/2$ with a velocity $p_{0}/m$,
where $m$ is the atomic mass, between $t$ and $t+\delta t$ are
those in the close vicinity of that border. The area of these zones
is simply $\delta p\, p_{0}\,\delta t/m$. The variation $\delta N_{x}$
in $N$ due to the velocity of the atoms can then be written:\begin{eqnarray}
\delta N_{x} & = & \frac{\delta p\, p_{0}\,\delta t}{m}\left[\rho\left(x_{0}-\frac{\delta x}{2},p_{0},t\right)-\rho\left(x_{0}+\frac{\delta x}{2},p_{0},t\right)\right]\nonumber \\
 & = & -\left(\delta x\,\delta p\,\delta t\right)\,\frac{p_{0}}{m}\,\frac{\partial}{\partial x}\rho\left(x_{0},p_{0},t\right)\label{eq:dNx}\end{eqnarray}

The last mechanism that changes $N$ is due to the changes in atomic
momentum, which is more tricky to evaluate. The atoms undergo cycles
where one photon is absorbed and another one is emitted. We consider
that all the underlying physics is described by a probability per
unit of time $\mathcal{P}\left(x,p,t,q\right)$: the probability to
change the atomic momentum by a quantity $q$ between $t$ and $t+\delta t$
is simply $\delta t\,\mathcal{P}\left(x,p,t,q\right)$. We assume
that the time interval $\delta t$ is small enough so that an atom
can undergo at most one photon scattering event. In such a case, $q=\hbar\left(k_{a}-k_{s}\right)$
where $k_{a}$ and $k_{s}$ are respectively the wavevector of the
absorbed and emitted photon. As a consequence, $q$ is bounded to
the range $\left[-2\hbar k,2\hbar k\right]$. A fraction, proportional
to $\delta t$, of the atoms contained in this region will change
their momentum by $q$. On the side $p_{+}=p_{0}+\delta p/2$, the
atom number variation $\delta N_{p+}$ is the difference between the
incoming atoms ($q<0)$ and the lost atoms ($q>0$): \begin{eqnarray}
\delta N_{p+} & = & \left(\delta x\,\delta t\right)\int_{0}^{2\hbar k}dq\int_{0}^{q}dq'\left(\mathcal{N}_{+}-\mathcal{N}_{-}\right)\label{eq:dNp1}\\
\mathrm{with\quad}\mathcal{N}_{\pm} & = & \mathcal{P}\left(x,p_{+}\pm q',t,\mp q\right)\,\rho\left(x,p_{+}\pm q',t\right)\nonumber \end{eqnarray}
To simplify this expression, we assume that the product $\mathcal{P}\left(x,p,t,q\right)\,\rho\left(x,p,t\right)$
varies slowly with $p$. Then, the integrand can be approximated by
its Taylor expansion in the vicinity of \foreignlanguage{english}{$p_{+}$}.
To recover the momentum diffusion process responsible for the non-zero
temperature of the trapped atoms, we have to expand the integrand
up to first order in $q'$. The inner integration is then straightforward
because the integrand becomes a linear function of $q'$. The same
calculation has to be done on the opposite side in $p_{-}=p_{0}-\delta p/2$,
which gives an analogous expression for $\delta N_{p-}$.

It appears natural to introduce the mean values of $q$ and $q^{2}$:
\begin{eqnarray}
F\left(x,p,t\right) & = & \int_{-2\hbar k}^{2\hbar k}\, q\, dq\,\mathcal{P}\left(x,p,t,q\right)\label{eq:Force}\\
D\left(x,p,t\right) & = & \int_{-2\hbar k}^{2\hbar k}\,\frac{q}{2}^{2}\, dq\,\mathcal{P}\left(x,p,t,q\right)\label{eq:Diffusion}\end{eqnarray}
These quantities can be interpreted as the mean force and the momentum
diffusion coefficient. The variation $\delta N_{p}=\delta N_{p+}+\delta N_{p-}$
in $N$ due to the change in momentum can be written as:\begin{equation}
\delta N_{p}=-\left(\delta x\,\delta p\,\delta t\right)\left[\frac{\partial}{\partial p}\left(F\rho\right)-\frac{\partial^{2}}{\partial p^{2}}\left(D\rho\right)\right]\label{eq:dNp}\end{equation}
Collecting all the contributions (\ref{eq:dNcoll}), (\ref{eq:dNx})
and (\ref{eq:dNp}) and going to the limit $\delta t\rightarrow0$,
we obtain the equation of evolution for the atomic density\begin{equation}
\frac{\partial}{\partial t}\rho+\frac{p}{m}\,\frac{\partial}{\partial x}\rho+\frac{\partial}{\partial p}\left(F\rho\right)-\frac{\partial^{2}}{\partial p^{2}}\left(D\rho\right)=-\frac{\rho}{\tau}+\lambda\label{eq:drhodt}\end{equation}

The three left terms of this equation are characteristic of a Vlasov
type kinetic model, except for the third term, as here the force depends
on the velocity of the particles. Thus this last term is a drift term,
and together with the fourth term, which is a relaxation term, they
denote a Fokker-Planck description. So the motion of cold atoms in
a MOT appears to be described by a Vlasov-Fokker-Planck (VFP) equation.
Thus MOTs are part of a large class of systems described by the VFP
equations, as e.g. plasmas \cite{plasma}, stars \cite{star}, granular
media \cite{benedetto2004} or electrons in a storage ring \cite{warnock2006}.
The different systems are characterized by the dependence of the force
$F$ on the phase space density $\rho$. For example, for a plasma
without magnetic fields, the Vlasov-Poisson-Fokker-Planck is used.

In these systems, the Fokker-Planck terms denote the immersion of
the particles in a thermal bath, which corresponds for the cold atoms
to the laser light. But in the case of cold atoms, a second bath,
namely the hot atoms of the residual gas, produces a second relaxation
term, as well as a source term (the right hand terms of Eq. \ref{eq:drhodt}):
indeed, the standard MOT is an open system, where the total population
can vary. This last point is crucial and cannot be neglected, as the
collision processes between hot atoms are the one and only source
of velocity redistribution allowing for a high MOT population. 

Because of the two relaxation processes involved in the MOTs, the
question of the mechanism in which originate the instabilities is
far from being trivial. Indeed, it is well known that the VFP equation
may have, for adequate parameters, unstable solutions. But as in \cite{distefano2003},
instabilities lead to variations of population, it is clear that the
loss and source terms can also generate unstable solutions. Therefore,
instabilities may originate in two different mechanisms, and it would
be interesting to search if one of them prevails. An interesting case
is that of the lower MOT in a double cell system. In this case, the
gas pressure in the cell is so low that the collision losses are negligible.
If the loading process by the upper MOT is stopped, Eq. \eqref{eq:drhodt}
becomes fully equivalent to plasmas equations, and only one relaxation
process remains. Thus it would be interesting to study experimentally
such a system, as the fact that instabilities persist or not in this
case could help in understanding the origin of these instabilities.

To go further, we need to evaluate the force \eqref{eq:Force} and
the diffusion coefficient \eqref{eq:Diffusion} for a specific atom.
To do so, we have to calculate the probability $\mathcal{P}\left(x,p,t,q\right)$
associated with the momentum exchange $q=\hbar\left(k_{a}-k_{s}\right)$.
However, as the probability to emit a spontaneous photon in the direction
$k_{s}$ is always identical to that for the opposite direction $-k_{s}$,
the contribution of spontaneous emission cancels in $F$, while in
$D$ the cross-term in $q^{2}$ vanishes. As a consequence, it is
not necessary to evaluate completely $\mathcal{P}\left(x,p,t,q\right)$,
but only the absorption rate for a photon of wavevector $k_{a}.$

\section{\label{sub:Radiative-force-and}Radiative forces and coefficients
of diffusion}

As stated above, we consider the simple scheme of a laser frequency
$\omega_{L}$ tuned in the vicinity of a $J=0\rightarrow J'=1$ transition
with a frequency $\omega_{0}$. These atoms are not suited for sub-Doppler
cooling mechanisms \cite{dalibard1989}, but these subtle processes
are not needed to describe the temperatures measured experimentally
in the cold atomic cloud obtained with high intensities and small
detunings for the trapping beams \cite{distefano2003}. In the case
of retroreflected trapping beams, this simple atom allows to develop
a model which convincingly reproduces the experimental observations
of instabilities, both deterministic \cite{distefano2004} and stochastic
\cite{hennequin2004}.

As usual in problems of laser cooling, we will extensively use the
fact that the time scales for the different processes are quite different.
The first time scale is given by the time a photon needs to go through
the sample. This time scale is so short that we can consider that
the light follows immediately any change in the sample. A second time
scale is given by the atomic response time $\Gamma^{-1}$, which is
in turn shorter than the third one related to the evolution of the
external degrees of freedom (the atomic position and velocity). This
inequality allows us to consider that the atomic internal state has
always reached its steady state. 

In the following, we consider an atom in $X$ with a velocity $V$,
where we use capital letters to distinguish the position and velocity
of this {}``real atom'' from the coordinates $\left(x,p\right)$
of a point in the phase space. The atoms are excited on a $J=0\rightarrow J'=1$
transition by two counterpropagating laser beams with opposite circular
polarizations. They also see a {}``bath'' of photons which are spontaneously
emitted by all the other atoms in the sample. These photons propagate
in all directions of the real $3D$ space and they have a broad spectrum.
Although it is not always true, we consider here that the effect of
the bath of spontaneous photons can be treated as a perturbation.
Then, we split the force and the diffusion coefficient into two parts:
one due to the trapping beams and the second one due to multiple scattering:
$F=F_{t}+F_{m}$ and $D=D_{t}+D_{m}$.

\subsection{Effect of the trapping beams}

The trap consists in two counterpropagating beams with the same frequency
$\omega_{L}$ and opposite polarizations (Fig. \ref{fig:Sketch}a).
The $\sigma^{-}$ beam has a wavevector $k_{-}=k\equiv\omega_{L}/c$
and has an intensity $I_{-}\left(X,t\right)$, while the $\sigma^{+}$
beam has a wavevector $k_{+}=-k$ and an intensity $I_{+}\left(X,t\right)$.
The frequency $\omega_{L}$ is slightly lower than the frequency $\omega_{0}$
of the atomic transition to insure an efficient Doppler cooling. The
sample is submitted to an inhomogeneous magnetic field aligned on
the propagation axis of the beams. For the sake of simplicity, we
assume that the magnetic field varies linearly: $B_{x}=bX$. The Zeeman
effect together with the cooling beams lead to a restoring force that
gathers the atoms in the vicinity of $X=0$. 

In the atomic rest frame, the apparent frequencies of the trapping
beams are Doppler shifted: $\omega_{\pm}=\omega_{L}\pm kV$. As the
two beams have opposite circular polarizations, the relevant detunings
are $\Delta_{\pm}=\Delta\pm\delta$, where we have introduced the
detuning $\Delta=\omega_{L}-\omega_{0}$ for an atom at rest with
no magnetic field, and the sum $\delta=kV+\mu_{B}bX/\hbar$ of the
Doppler shift $kV$ and the Zeeman shift $\mu_{B}B_{x}/\hbar$. As
the two beams are circularly polarized, the excited state $\left|0\right\rangle $
plays no role in the present configuration. The coupling between the
ground state $\left|g\right\rangle $ and the excited states $\left|\pm\right\rangle $
is described by the Rabi frequencies $\left|\Omega_{\pm}\right|=\Gamma\sqrt{I_{\pm}/2I_{s}}$,
where $I_{s}$ is the saturation intensity. It is then possible to
find exactly the steady state of the density matrix, by solving the
master equation. We are not interested here in the explicit form of
the populations or of the coherences, which are given in a previous
article \cite{hennequin2004}. The only important point is that the
stationary populations $\Pi_{\pm}$ in the excited states $\left|\pm\right\rangle $
can be written as a function of the relevant parameters $\Delta$,
$\delta$, $I_{+}$ and $I_{-}$. Note that these stationary populations
do not depend on the position or the velocity of the atom explicitly,
but only through $\delta$, $I_{+}$ and $I_{-}$. 

The key point is that the simplicity of the atomic structure means
that $\gamma_{\pm}=\Gamma\Pi_{\pm}$ are the emission rates of spontaneous
$\sigma^{\pm}$ polarized photons. The exact expressions of $\Pi_{\pm}$
can be found in \cite{hennequin2004}. Again because of the simple
$J=0\rightarrow J'=1$ transition, a photon with the same polarization
has to be absorbed before each spontaneous emission process. As a
consequence, we obtain a very simple expression for the radiative
force due to the trapping beams, together with the expression for
the diffusion coefficient due to the trapping beams:\begin{subequations}\label{eq:ftdt}\begin{eqnarray}
F_{t} & = & \hbar k\left(\gamma_{-}-\gamma_{+}\right)\label{eq:ft}\\
D_{t} & = & \frac{7}{5}\hbar^{2}k^{2}\left(\gamma_{-}+\gamma_{+}\right)\label{eq:dt}\end{eqnarray}
 \end{subequations}where a contribution $2k^{2}/5$ in $D_{t}$ comes
from the mean value of the square of the projected component of the
spontaneous wavevector. To evaluate this average, we consider the
diagram of emission of a circular electric dipole ($3/10$ in the
plane and $2/5$ in the orthogonal direction).

\subsection{Effect of multiple scattering}

The contribution of multiple scattering is trickier to evaluate. We
have to consider our atom in a bath of incoherent photons which propagate
in all directions. Because of their physical origin, photons do not
differ in spectrum wherever the scatterer is and whatever their direction
of propagation is. Therefore, the rate of absorption of photons going
in a given direction is simply proportional to the flux of photons
that travel in this direction. The contribution to that flux of photons
scattered just once is easy to evaluate. The flux of photons scattered
more than once is more difficult to estimate, but it is in general
much smaller than the previous one, and we neglect it in the following.
The flux $\Psi_{+}$(resp. $\Psi_{-}$) of photons, travelling to
the right (resp. to the left) and impinging on an atom in $X$, is
just half of the total number of photons scattered by the atoms on
the left (resp. on the right) of $X$.\begin{subequations}\label{eq:Psi}
\begin{eqnarray}
\Psi_{+}\left(X,t\right) & = & \frac{1}{2}\int_{-\infty}^{X}dx'\int dp\,\rho\left(x',p,t\right)\,\left(\gamma_{-}+\gamma_{+}\right)\label{eq:Psi+}\\
\Psi_{-}\left(X,t\right) & = & \frac{1}{2}\int_{X}^{+\infty}dx'\int dp\,\rho\left(x',p,t\right)\,\left(\gamma_{-}+\gamma_{+}\right)\label{eq:Psi-}\end{eqnarray}
\end{subequations}In these expressions, we do not differentiate the
polarization of the scattered photons. We just sum up the contribution
$\gamma_{+}$ of the mainly $\sigma^{+}$ polarized photons and the
contribution $\gamma_{-}$ of the mainly $\sigma^{-}$ polarized photons.
It is interesting to note that the sum $\Psi_{+}+\Psi_{-}$ of these
two fluxes does not depend on $X$. We shall see in the next section
that these fluxes have much simpler expressions.

To evaluate the force exerted by the scattered photons on the atoms,
we have to know the fraction of photons which are absorbed and the
momentum carried by each photon. This momentum is smaller than $\hbar k$,
because we have to keep here only the on-axis component for photons
travelling in all directions of the real $3D$ space. Considering
the emission diagram of a circular dipole, we obtain $\left\langle k_{a}\right\rangle =9k/16$
and $\left\langle k_{a}^{2}\right\rangle =2k^{2}/5$. This leads to
the following expressions for $F_{m}$ and $D_{m}$:\begin{subequations}\label{eq:fmdm}\begin{eqnarray}
F_{m} & = & \sigma_{R}\,\frac{9}{16}\,\hbar k\left(\Psi_{+}-\Psi_{-}\right)\\
D_{m} & = & \sigma_{R}\,\frac{4}{5}\,\hbar^{2}k^{2}\left(\Psi_{-}+\Psi_{+}\right)\end{eqnarray}
\end{subequations}where we have introduced re-absorption cross-section
$\sigma_{R}$.

In the limit of very low intensities ($\Omega\ll\Gamma$), we can
give an estimate of $\sigma_{R}$. In this case, it is well-known
that for a 2-level atom, the emission spectrum is mainly elastic \cite{cohen}.
This result holds for the $J=0\rightarrow J'=1$ transition considered
here and the scattered photons have the same frequency $\omega_{L}$
as the trapping lasers. To be rigorous, Doppler broadening should
be taken into account, as the scatters are moving and the emitted
photons do not propagate in the direction of the trapping beams. However,
we neglect this broadening here, because it is smaller than both the
detuning and the natural linewidth. We also neglect the Zeeman shift
and the Doppler shift for the absorber. The absorption cross-section
is thus reduced by a factor $\Gamma^{2}/\left(4\Delta^{2}+\Gamma^{2}\right)$
with respect to the cross-section at resonance $\sigma_{0}=3\lambda^{2}/2\pi$
. In the very low intensity limit, we have:\begin{eqnarray*}
\sigma_{R} & =\sigma_{0} & \frac{\Gamma^{2}}{4\Delta^{2}+\Gamma^{2}}\end{eqnarray*}

For higher but still modest intensities, a new contribution shows
up : the blue sideband of the Mollow triplet \cite{mollow}, which
excites resonantly the atomic transition, grows up as $\Omega^{2}/\left(4\Delta^{2}+\Gamma^{2}\right)$.
As soon as $\Omega\gg\Gamma$, this new resonant contribution dominates,
even in the low saturation regime ($1\gg\Omega^{2}/\left(4\Delta^{2}+\Gamma^{2}\right)$).

For even higher intensities, as we know the exact steady state of
the density matrix, we can, in principle, compute the spectrum of
the scattered light and the absorption cross-section, following what
is usually done for 2-level atoms \cite{pruvost2000}. However, contrary
to the previous publications \cite{pruvost2000,sesko1991}, we do
not evaluate $\sigma_{R}$ for a 2-level atom, to remain consistent
with the model of a $J=0\rightarrow J'=1$ transition where at least
one of the excited state sub-levels does not interact with the trapping
beams. As the exact calculation is quite heavy, we shall not make
here the full calculation in the general case, but we shall restrict
ourselves to the simpler situation considered in section \ref{sec:Approximations}.

\section{Propagation of the trapping beams\label{sub:Propagation}}

In this section, we derive the equation of propagation for the trapping
beams assuming that we know the atomic density $\rho\left(x,p,t\right)$.
We are not interested in the phase of the laser beams, but just in
their intensities. We have seen in the previous section that the rates
$\gamma_{\pm}=\Gamma\Pi_{\pm}$ are the absorption rates of $\sigma^{\pm}$
polarized photons for an atom in $X$ moving with velocity $V$. To
know how many photons are absorbed between $x$ and $x+\delta x$,
we just have to sum the contributions of all atoms in $X=x$ with
all possible velocities $V=p/m$. These photons are taken from the
trapping beam with the appropriate polarization, so we get :\begin{subequations}\label{eq:dl}\begin{eqnarray}
\frac{\partial I_{+}}{\partial x} & = & +\hbar\omega_{L}\int_{-\infty}^{+\infty}dp\,\rho\left(x,p,t\right)\,\gamma_{+}\label{eq:dI+}\\
\frac{\partial I_{-}}{\partial x} & = & -\hbar\omega_{L}\int_{-\infty}^{+\infty}dp\,\rho\left(x,p,t\right)\,\gamma_{-}\label{eq:dI-}\end{eqnarray}
\end{subequations} The plus sign in (\ref{eq:dI+}) comes from the
fact that the $\sigma^{+}$ polarized beam propagates backwards. So
$I_{+}\left(x,t\right)$ increases with $x$, while $I_{-}\left(x,t\right)$
decreases.

Plugging (\ref{eq:dl}) in the fluxes (\ref{eq:Psi}) and performing
formally the integration on positions, we get :\begin{subequations}\begin{eqnarray*}
\Psi_{+}\left(x,t\right) & = & \frac{I_{+}\left(x,t\right)-I_{-}\left(x,t\right)-I_{+}\left(-\infty,t\right)+I_{-}\left(-\infty,t\right)}{2\hbar\omega_{L}}\\
\Psi_{-}\left(x,t\right) & = & \frac{I_{+}\left(\infty,t\right)-I_{-}\left(\infty,t\right)-I_{+}\left(x,t\right)+I_{-}\left(x,t\right)}{2\hbar\omega_{L}}\end{eqnarray*}
\end{subequations}The intensities $I_{-}\left(-\infty,t\right)$
and $I_{+}\left(\infty,t\right)$ are the incoming laser intensities,
which are assumed to be constant. As soon as we know the atomic phase
space density $\rho\left(x,p,t\right)$, we can solve the equations
\eqref{eq:dl} to get the spatial evolution of the intensities. However,
the evolution of the phase space density depends on the laser intensities
through the radiative forces and the diffusion coefficients. Thus,
the equations \eqref{eq:drhodt} and \eqref{eq:dl} form a set of
coupled equations which have to be solved together.

It is important to note that Eqs \eqref{eq:dl} introduce in the force
a term which, in general, is not present in other systems described
by the VFP equation. This difference has to be taken into account
if we want to apply results obtained for plasmas to the dynamics of
MOTs. Indeed, in plasmas, the particles do not act, by definition,
on the thermal bath. On the contrary, in MOTs, atoms act on the beams
(the equivalent of the thermal bath, as seen above), through the absorption
and the so-called shadow effect. Thus it leads to another indirect
dependence of the force on the density.

\section{Approximation for the atomic response\label{sec:Approximations}}

In the general case, the equations \eqref{eq:drhodt} and \eqref{eq:dl}
describing the MOT are highly non linear, and have no simple solution.
Numerical simulations could give some solutions, but they could not
give any physical insight of the phenomenon. However, starting from
this general model, we can now consider some approximations allowing
going further in the understanding of the MOT dynamics. Through these
approximations, we aim to determine the expressions \eqref{eq:ftdt}
of the trapping beam contribution and \eqref{eq:fmdm} of the multiple
scattering contribution to the mean force and diffusion coefficient
of the MOT.

\subsection{Trapping forces}

Concerning the trapping beam components, Eqs \eqref{eq:ftdt} shows
that we need essentially to explicit the scattering rates $\gamma_{\pm}=\Gamma\,\Pi_{\pm}$,
introduced as the emission rates of spontaneous $\sigma^{\pm}$ polarized
photons, and also as the absorption rates of the corresponding trapping
photons. In the case of a $J=0\rightarrow J'=1$ transition, these
rates are exactly computable, as in \cite{hennequin2004}. However,
the obtained expressions are very heavy and difficult to interpret.
To have a better physical insight, we will simplify these expressions
through approximations valid in our domain of interest. 

The usual approach consists in considering the low saturation limit.
In this case, each wave acts on the atoms independently, and the forces
and the diffusion coefficient can be computed. But experiments are
seldom realized in this low intensity limit. For slightly higher intensities,
the calculation of the next order of perturbation (fourth order in
field) must be considered. But this calculation is not straightforward.
Indeed, the third order of perturbation builds up Zeeman coherence
between the excited sublevels, which lead to non-trivial modifications
of the populations at the fourth order. The calculation can be done,
but the cross-terms do not allow any simple interpretation. Moreover,
to evaluate the effect of multiple scattering, we need to know the
absorption cross section for the bath of photons that have been scattered
elsewhere. It requires one more step of perturbation by an extra weak
field, which is not easy to implement.

This first approximation is usually followed by another approximation,
namely the low velocity limit, where the total trapping force is linearized
in $V$. It naturally splits in a term due to the intensity imbalance
and in a friction term. In the same way, the gradient of the magnetic
field introduces a restoring force, linear in $X$. The condition
of validity for this approximation is $\left|\delta\right|\ll max\left(\left|\Delta\right|,\Gamma\right)$,
where $\delta$ is the sum of the Doppler and Zeeman shifts, as introduced
in section \ref{sub:Radiative-force-and}.

Let us consider here another approach, where the approximation $\delta$
is applied before going to the low saturation limit. The total shift
$\delta$ is considered as a perturbation, that is expanded to first
order only. At a first glance, this choice seems quite surprising,
as the {}``perturbation'' is diagonal in the natural basis $\left\{ \left|g\right\rangle ,\left|+\right\rangle ,\left|-\right\rangle \right\} $.
However, the good basis to do the calculation is not the natural one,
and then the shift $\delta$ leads to off-diagonal terms.

At zeroth order, the case $\delta=0$ corresponds to an atom interacting
with a field which has both $\sigma^{\pm}$ components. It is thus
natural to introduce the coupled $\left|C\right\rangle $ and non-coupled
$\left|N\right\rangle $ states: \begin{eqnarray*}
\left|C\right\rangle  & = & \frac{\Omega_{+}\left|+\right\rangle +\,\Omega_{-}\left|-\right\rangle }{\Omega}\\
\left|N\right\rangle  & = & \frac{\Omega_{-}^{*}\left|+\right\rangle -\,\Omega_{+}^{*}\left|-\right\rangle }{\Omega}\end{eqnarray*}
where $\Omega_{\pm}$ are the generalized (C-number) Rabi frequencies
that include the complex phase of the fields and $\Omega=\sqrt{\left|\Omega_{+}\right|^{2}+\left|\Omega_{-}\right|^{2}}$
describes the coupling between the ground state and the coupled one.
Then the problem reduces to that of a 2-level atom interacting with
a single laser field. The steady state population of the coupled state
is simply:\[
\Pi_{C}^{(0)}=\frac{\Omega^{2}}{4\Delta^{2}+2\Omega^{2}+\Gamma^{2}}\]
 which gives the value of the scattering rates :\[
\gamma_{\pm}^{(0)}=\Gamma\,\frac{\Omega_{\pm}^{2}}{4\Delta^{2}+2\Omega^{2}+\Gamma^{2}}\]
The radiative force due to the beam imbalance and the coefficient
of diffusion follows:\begin{eqnarray*}
F_{t}^{(0)} & = & \frac{\hbar\, k\,\Gamma}{4\Delta^{2}+2\Omega^{2}+\Gamma^{2}}\left(\left|\Omega_{-}\right|^{2}-\left|\Omega_{+}\right|^{2}\right)\\
D_{t}^{(0)} & = & \frac{7}{5}\,\hbar^{2}k^{2}\,\Gamma\frac{\Omega^{2}}{4\Delta^{2}+2\Omega^{2}+\Gamma^{2}}\end{eqnarray*}
It is interesting to note that, in these expressions, we get the same
result if we replace, in the expressions obtained in the low saturation
regime, the usual denominator $\left(4\Delta^{2}+\Gamma^{2}\right)$
by $\left(4\Delta^{2}+2\Omega^{2}+\Gamma^{2}\right)$.

At first order, when $\delta\neq0$, we introduce the same coupled
and non-coupled states as above, and a first order perturbation in
$\delta$ gives the friction and restoring forces. The diagonal (in
the natural basis) part of the hamiltonian $H_{A}$ writes: \begin{eqnarray*}
H_{A} & = & -\left(\Delta_{+}\left|+\right\rangle \left\langle +\right|+\,\Delta_{-}\left|-\right\rangle \left\langle -\right|\right)\\
 & = & -\Delta_{N}\left|N\right\rangle \left\langle N\right|-\,\Delta_{C}\left|C\right\rangle \left\langle C\right|\\
 &  & -2\frac{\delta}{\Omega^{2}}\left(\Omega_{+}\Omega_{-}\left|N\right\rangle \left\langle C\right|+\Omega_{+}^{*}\Omega_{-}^{*}\left|C\right\rangle \left\langle N\right|\right)\end{eqnarray*}
with \begin{eqnarray*}
\Delta_{C} & = & \Delta+\delta\frac{\left|\Omega_{+}\right|^{2}-\left|\Omega_{-}\right|^{2}}{\Omega^{2}}\\
\Delta_{N} & = & \Delta-\delta\frac{\left|\Omega_{+}\right|^{2}-\left|\Omega_{-}\right|^{2}}{\Omega^{2}}\end{eqnarray*}
The slight change from $\Delta$ to $\Delta_{C}$ introduces a first
order correction in the population of the excited coupled state.\begin{equation}
\Pi_{C}^{(1)}=\frac{8\,\Delta\,\delta\left(\left|\Omega_{-}\right|^{2}-\left|\Omega_{+}\right|^{2}\right)}{\left(4\Delta^{2}+2\Omega^{2}+\Gamma^{2}\right)^{2}}\label{eq:pop_C1}\end{equation}
The off-diagonal perturbation terms induce, at the first order, coherences
between the non-coupled state and the two other states. But only the
coherences in the excited state are needed to evaluate the populations
$\Pi_{\pm}$\begin{eqnarray}
\left\langle N\left|\sigma^{(1)}\right|C\right\rangle  & = & \frac{-2\,\delta\,\Omega_{+}\Omega_{-}}{4\Delta^{2}+2\Omega^{2}+\Gamma^{2}}\,\frac{4\,\Gamma}{4\Gamma\Delta-i\,\left(2\Gamma^{2}+\Omega^{2}\right)}\label{eq:coh_NC1}\\
\left\langle C\left|\sigma^{(1)}\right|N\right\rangle  & = & \left\langle N\left|\sigma^{(1)}\right|C\right\rangle ^{*}\label{eq:coh_CN1}\end{eqnarray}
The correction in the scattering rates is deduced from the expressions
of the population \eqref{eq:pop_C1} and of the coherences \eqref{eq:coh_NC1}
and \eqref{eq:coh_CN1}:\begin{eqnarray*}
\gamma_{\pm} & = & \frac{\mp8\,\Delta\,\delta\,\Gamma}{\left(4\Delta^{2}+2\Omega^{2}+\Gamma^{2}\right)^{2}}\\
 &  & \times\left(\left|\Omega_{\pm}\right|^{2}+2\left|\Omega_{-}\right|^{2}\left|\Omega_{+}\right|^{2}\frac{4\Gamma^{2}-\Omega^{2}}{16\Gamma^{2}\Delta^{2}+\left(2\Gamma^{2}+\Omega^{2}\right)^{2}}\right)\end{eqnarray*}
 The evaluation of the friction force is then straightforward: \begin{eqnarray*}
F_{t}^{(1)} & = & \frac{\hbar\, k\,\Gamma\,8\,\Delta\,\delta}{\left(4\Delta^{2}+2\Omega^{2}+\Gamma^{2}\right)^{2}}\\
 &  & \times\left(\Omega^{2}+4\left|\Omega_{-}\right|^{2}\left|\Omega_{+}\right|^{2}\frac{4\Gamma^{2}-\Omega^{2}}{16\Gamma^{2}\Delta^{2}+\left(2\Gamma^{2}+\Omega^{2}\right)^{2}}\right)\\
D_{t}^{(1)} & = & \frac{7}{5}\,\hbar^{2}k^{2}\,\Gamma\frac{8\,\Delta\,\delta\left(\left|\Omega_{-}\right|^{2}-\left|\Omega_{+}\right|^{2}\right)}{\left(4\Delta^{2}+2\Omega^{2}+\Gamma^{2}\right)^{2}}\end{eqnarray*}
It is easy to check that the force calculated in this way is rather
a friction force, as soon as $\Delta<0$. The first term of the parenthesis,
$\Omega^{2}$, together with the prefactor, corresponds to what is
guessed from the unsaturated expression: the original denominator
$\left(4\Delta^{2}+\Gamma^{2}\right)$ becomes $\left(4\Delta^{2}+2\Omega^{2}+\Gamma^{2}\right)$.
However, a second term is needed to take into account properly the
cross-saturation effects. This term is often forgotten by the authors,
as for instance in \cite{pohl2006}, whereas it can modify significantly
the spring constant of the trap, as it is of the same order of magnitude
as $\Omega^{2}$. In particular, the minus sign means that, depending
on the parameters, the spring constant is increased or decreased significantly.
On the contrary, the correction $D_{t}^{(1)}$ to the coefficient
of diffusion is usually very small, because $\left|\Omega_{-}\right|^{2}\simeq\left|\Omega_{+}\right|^{2}$.

\subsection{Multiple scattering}

To evaluate the effect of multiple scattering, we neglect, as usual,
the Doppler broadening due to the motion of the emitter and the Doppler
and Zeeman shifts for the re-absorber. Thus, we consider that $\delta$
vanishes. Eqs \eqref{eq:fmdm} shows that we have to evaluate the
re-absorption cross-section $\sigma_{R}$, taking into account the
fluorescence spectrum and the absorption spectrum, which are affected
by the intense trapping field.

When the laser field is intense enough, the dressed atom in the secular
limit allows a simple evaluation of the fluorescence spectrum and
of the re-absorption cross-section \cite{pruvost2000,cohen}. However,
instead of considering a 2-level atom, we remain consistent with the
model of a $J=0\rightarrow J'=1$ transition where we have three sublevels
in the excited state. We have also to consider that the polarization
of the photon scattered by an atom somewhere in the sample will not
match exactly the polarization of the local field. For the sake of
simplicity, let us consider here the case where the trapping beams
have the same intensity, resulting in a linear polarization for the
total trapping field. The consequence of the polarization mismatch
is that, on average, one half of the scattered photons has the polarization
of the local field while the other half has the orthogonal polarization.
The detailed calculation is done in the appendix. The energy levels
of the dressed atom are sketched in Fig. \ref{fig:At_hab}. In Fig.
\ref{fig:At_hab}a, we have evidenced the fluorescence transitions
that lead to the Mollow triplet. On the other hand, the absorption
lines are drawn on Fig. \ref{fig:At_hab}b, considering all the possible
polarizations for the incident photon. Schematically, we have to consider
the overlap of the four components of the Mollow triplet with the
four absorption lines of the 3-level atom, as represented in Fig.
\ref{fig:At_hab}c. Some care has to be taken to compare the various
contributions in the different regimes. Let us now examine the four
main situations:

\begin{figure}
\includegraphics[width=8cm]{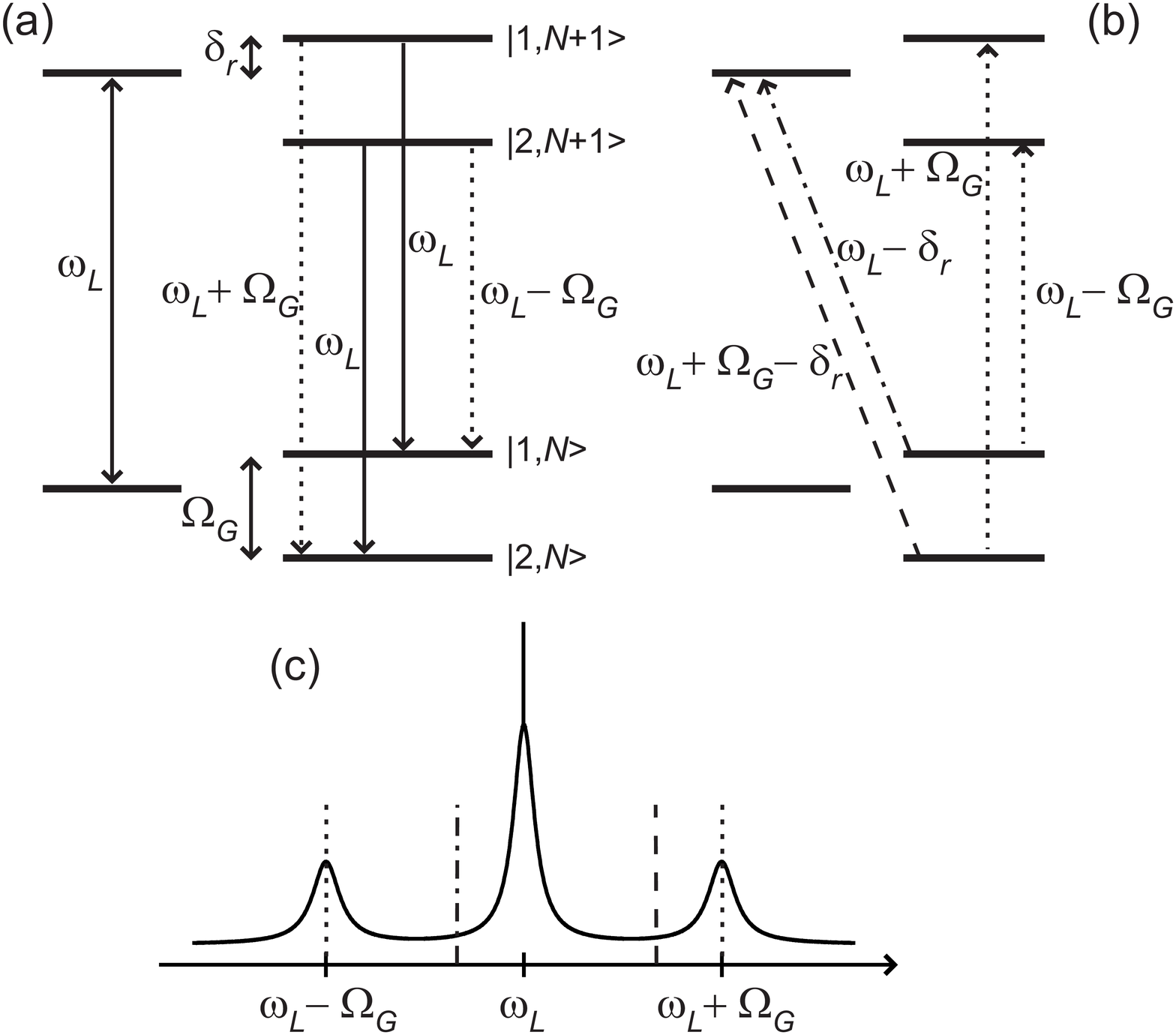}\caption{\label{fig:At_hab}Sketch of the energy levels of the dressed-atom.
a) Spontaneous emission lines leading to the Mollow triplet. b) Absorption
lines. c) Overlap of the fluorescence spectrum (solid line) with the
absorption lines of the 2-level system (dotted lines) and with those
due to the third level (dashed lines).}

\end{figure}

(i) $\Omega,\left|\Delta\right|\gg\Gamma$. In general, for large
detunings and intensities and large light-shifts ($\delta_{r}=\frac{1}{2}\left(\sqrt{\Delta^{2}+\Omega^{2}}-\left|\Delta\right|\right)\gg\Gamma$),
the resonant term of the 2-level system dominates the sum, and all
the other terms should be dropped to remain consistent with the secular
approximation done before. The resulting re-absorption cross-section
has thus the following simple expression:\begin{equation}
\sigma_{R}=\frac{\sigma_{0}}{4}\,\frac{\Delta^{2}\Omega^{2}}{4\Delta^{4}+8\Delta^{2}\Omega^{2}+3\Omega^{4}}\label{eq:eta_resonant}\end{equation}
This result is consistent, within a factor of 2, with the limit values
given for $\Omega\gg\left|\Delta\right|\gg\Gamma$ and $\left|\Delta\right|\gg\Omega\gg\Gamma$
in the appendix of Ref. \cite{pruvost2000}. 

(ii) $\left|\Delta\right|\gg\Omega$. The contribution due to the
non-coupled level is never exactly resonant. However, in this case,
the light-shift, $\delta_{r}\simeq\Omega^{2}/4\left|\Delta\right|$,
can be arbitrarily small, leading to an almost resonant behaviour
if $\delta_{r}\lesssim\Gamma$. In the case where $\left|\Delta\right|\gg\Omega\gg\Gamma$
and $\Gamma\gg\Omega^{2}/4\left|\Delta\right|$, we take into account
the quasi resonant contribution of the third level:\begin{equation}
\sigma_{R}=\frac{\sigma_{0}}{8}\,\frac{\Omega^{2}}{\Delta^{2}}\label{eq:eta _quasiresonant}\end{equation}
Note that \eqref{eq:eta _quasiresonant} is not the limit of \eqref{eq:eta_resonant}
when $\left|\Delta\right|\gg\Omega$, although it has the same form.

(iii) $\Omega\ll\Gamma$. When the intensity is very small, the resonant
or quasi-resonant contributions go to zero and the leading term corresponds
to the reabsorption of the elastic component:\begin{equation}
\sigma_{R}=\frac{\sigma_{0}}{4}\,\frac{\Gamma^{2}}{\Delta^{2}}\label{eq:eta_faible_int}\end{equation}

(iv) $\left|\Delta\right|\ll\Gamma$. For small detunings, the reabsorption
by a 2-level atom goes to zero because the transition is strongly
saturated. Then, the contribution of the third level, although it
is non-resonant, has to be taken into account. We have:\begin{equation}
\sigma_{R}=\sigma_{0}\frac{25}{48}\,\frac{\Gamma^{2}}{\Omega^{2}}\label{eq:eta_resonance}\end{equation}

The above results allow us to write in the different regimes the contribution
of the multiple scattering to the MOT dynamics. These expressions
are rather simple as compared to those found in the literature. It
is interesting to note that $\sigma_{R}$ is always larger than the
absorption cross-section for the trapping beams. This is very often
assumed in previous papers, as stated in \cite{sesko1991}. We have
demonstrated here that, in the secular regime for three level atoms,
it is always true. The above expressions of $\sigma_{R}$, together
with the expression of the trapping force derived in the previous
paragraph, allow us to write a full model of the MOT, which can be
used as a basis for future analyses.

\section{Conclusion}

In this paper, we built a model for a 1D Magneto-Optical Trap as general
as possible. We show that such a trap is described by a Vlasov-Fokker-Planck
equation with a second relaxation term and a source term, both originating
in the bath of hot atoms of the atomic vapor. This VFP equation is
coupled to a set of two differential equations describing the beam
propagation in the cold atoms. This system could be considered as
relatively similar to plasmas, where the role of the thermal bath
is played by the trapping beams. However, it appears that the MOT
differs from plasmas on two important points: the second thermal bath,
formed by the hot atoms, induces new interactions as compared to plasmas;
the trapping beams are not a {}``thermal'' bath, as the atoms act
on them through the absorption. 

In the last section of this paper, we derive in a more detailed way
the equations established in the previous sections, and compare the
results with those found in the litterature. We show that important
correction terms must be taken into account in the evaluation of the
trapping forces, and we establish the expression of the multiple scattering
in different situations.

These equations aimed to be a basis for future developments. They
should contribute to obtain a better agreement between the experimental
observations and the theoretical predictions. Let us remember that
the previous existing models were only in qualitative agreement with
the experimental observations, in particular for the situations out
of equilibrium. Using this model to study the instabilities in the
cold atoms should give a better insight of the mechanism in which
they originate. Thus a natural continuation of the present work would
be to extend this model to the 3D traps.

\section{Appendix}

In this appendix, we present the detailed calculations of the re-absorption
cross section, resulting from the overlap of the fluorescence spectrum
of an atom somewhere in the sample with the absorption spectrum of
another atom elsewhere. In the $J=0\rightarrow J'=1$ transition considered
here, we have three sublevels in the excited state, and the polarization
of the scattered photon will not match exactly the polarization of
the local field. For the sake of simplicity, let us consider here
the case where the trapping beams have the same intensity. In real
experiments, it cannot be true everywhere in the sample because of
the shadow effect, but the deviation remains on the order of 10\%.
In this simple case, the resulting polarization of the total trapping
field is everywhere linear, but its direction follows an helix. The
laser wavelength is the scale for changing the relative phase between
$\Omega_{+}$ and $\Omega_{-}$, because the two beams are contrapropagating.
As soon as we are not interested in what happens at length scales
smaller than the laser wavelength, we can estimate that one half of
the photons in the bath will interact with the transition between
the ground state and the coupled one, while the other half of them
can excite the atom to the non-coupled level. The total re-absorption
cross-section, $\sigma_{R}$, is thus the average of the usual cross-section,
$\sigma_{R}^{(2)}$, calculated with a 2-level atom \cite{pruvost2000}
and of a new contribution, $\sigma_{R}^{(3)}$, coming from the presence
of an excited level which is not coupled to the trapping field. The
empty non-coupled state allows a strong absorption of the scattered
light, while the absorption by the coupled state can be saturated
by the trapping field. 

As usual, when one works with the dressed atom model, the calculations
are much simpler in the secular approximation. This approximation
is valid as soon as the energy splitting is larger than the typical
width of the levels, which is expressed by the condition: \begin{equation}
\sqrt{\Delta^{2}+\Omega^{2}}\gg\Gamma\label{eq:seculaire}\end{equation}
Please note that this relation does not require that both the detuning
and the intensity are large, and some interesting limits can be examined
with either $\Omega\lesssim\Gamma$ or $\left|\Delta\right|\lesssim\Gamma$,
in the secular limit. First, we calculate the emission spectral density,
$S\left(\omega\right)$, for an atom illuminated with the two trapping
beams. Then, we evaluate the absorption spectra of another atom, also
interacting with the trapping field. Two contributions appear in the
absorption process: on one hand, the coupled excited level can still
absorb light, leading to a contribution which is the one of a 2-level
atom, $\sigma_{A}^{(2)}\left(\omega\right)$, and on the other hand,
the non-coupled level has to contribute with $\sigma_{A}^{(3)}\left(\omega\right)$.
Finally, we evaluate the total re-absorption cross-section, $\sigma_{R}$,
with:\begin{eqnarray}
\sigma_{R} & = & \frac{1}{2}\left(\sigma_{R}^{(2)}+\sigma_{R}^{(3)}\right)\label{eq:sigma_R}\\
\sigma_{R}^{(2)} & = & \int\,\sigma_{A}^{(2)}\left(\omega\right)\, S\left(\omega\right)\, d\omega\nonumber \\
\sigma_{R}^{(3)} & = & \int\,\sigma_{A}^{(3)}\left(\omega\right)\, S\left(\omega\right)\, d\omega\nonumber \end{eqnarray}
All these calculations are done using the formalism of the dressed
atom, in the secular limit (\ref{eq:seculaire}). The normalized fluorescence
spectrum is given by:\begin{eqnarray*}
S\left(\omega\right) & = & \frac{\left(c^{2}-s^{2}\right)^{2}}{c^{4}+s^{4}}\delta\left(\omega-\omega_{L}\right)+\frac{\left(2c^{2}s^{2}\right)^{2}}{c^{4}+s^{4}}L\left(\omega-\omega_{L},\Gamma_{p}\right)\\
 &  & +c^{2}s^{2}L\left(\omega-\omega_{L}-\Omega_{G},\Gamma_{c}\right)\\
 &  & +c^{2}s^{2}L\left(\omega-\omega_{L}+\Omega_{G},\Gamma_{c}\right)\end{eqnarray*}
where $\Omega_{G}=\sqrt{\Delta^{2}+\Omega^{2}}$ is the generalized
Rabi frequency. The transformation from the bare basis to the dressed
one is given by $c=cos\left(\theta\right)$, $s=sin\left(\theta\right)$,
with the angle $\theta$ defined by $tg\left(2\theta\right)=-\frac{\Omega}{\Delta}$.
$\delta\left(\omega-\omega_{L}\right)$ is the Dirac delta function,
and $L\left(\omega,\Gamma\right)$ is the normalized Lorentzian:\[
L\left(\omega,\Gamma\right)=\frac{1}{\pi}\,\frac{\Gamma}{\omega^{2}+\Gamma^{2}}\]
and the relaxation rates for the populations and the coherences are:\begin{eqnarray}
\Gamma_{p} & = & \Gamma\left(1-2c^{2}s^{2}\right)\label{eq:Gamma_pop}\\
\Gamma_{c} & = & \frac{\Gamma}{2}\left(1+2c^{2}s^{2}\right)\label{eq:Gamma_coh}\end{eqnarray}
The first term in $S\left(\omega\right)$ corresponds to the elastic
scattering, while the three other terms are inelastic components.
The last two terms are the sidebands of the Mollow triplet, centered
in $\omega_{L}\pm\Omega_{G}$, which are proportional to the intensity
when $\left|\Delta\right|\gg\Omega$.

The absorption spectra are given by:\begin{eqnarray*}
\sigma_{A}^{(2)}\left(\omega\right) & = & \sigma_{0}\frac{\pi\Gamma}{2}\frac{c^{2}-s^{2}}{c^{4}+s^{4}}c^{4}L\left(\omega-\omega_{L}-\Omega_{G},\Gamma_{c}\right)\\
 &  & -\sigma_{0}\frac{\pi\Gamma}{2}\frac{c^{2}-s^{2}}{c^{4}+s^{4}}s^{4}L\left(\omega-\omega_{L}+\Omega_{G},\Gamma_{c}\right)\\
\sigma_{A}^{(3)}\left(\omega\right) & = & \sigma_{0}\frac{\pi\Gamma}{2}c^{6}L\left(\omega-\omega_{L}-\Omega_{G}+\delta_{r},\Gamma_{2}\right)\\
 &  & +\sigma_{0}\frac{\pi\Gamma}{2}s^{6}L\left(\omega-\omega_{L}+\delta_{r},\Gamma_{1}\right)\end{eqnarray*}
where $\delta_{r}$ is the light-shift, and $\Gamma_{1,2}$ are the
relaxation rates for the coherences between the non-coupled level
and the dressed states. \begin{eqnarray}
\delta_{r} & = & \frac{1}{2}\left(\Delta+\sqrt{\Delta^{2}+\Omega^{2}}\right)\label{eq:depl_lum}\\
\Gamma_{1} & = & \frac{\Gamma}{2}\left(1+c^{2}\right)\label{eq:Gamma1}\\
\Gamma_{2} & = & \frac{\Gamma}{2}\left(1+s^{2}\right)\label{eq:Gamma2}\end{eqnarray}
The absorption spectrum of the 2-level part consists in two lines
centred in $\omega_{L}\pm\Omega_{G}$, which means that they are resonantly
excited by the two sidebands of the Mollow triplet. The two absorption
lines due to the third level are centred in $\omega_{L}+\Omega_{G}-\delta_{r}$
and in $\omega_{L}-\delta_{r}$, then the excitation by the fluorescence
light is not resonant if $\delta_{r}\gg\Gamma$. Because of the secular
approximation we made for the calculation, it is incorrect to write
$\sigma_{R}$ as the sum of 16 terms, coming from the overlap of the
four components of the fluorescence with the four absorption lines.
We have to consider separately the different cases.

In general, for large detunings and intensities ($\Omega,\left|\Delta\right|\gg\Gamma$)
and large light-shifts ($\delta_{r}\gg\Gamma$), the resonant term
of the 2-level system dominates the sum, and one should drop all the
other terms to be consistent with the secular approximation done before.
The resulting cross-section has thus the following simple expression:\begin{equation}
\sigma_{R}=\frac{\sigma_{0}}{4}\,\frac{\Delta^{2}\Omega^{2}}{4\Delta^{4}+8\Delta^{2}\Omega^{2}+3\Omega^{4}}\label{eq:resonant}\end{equation}
As already mentioned, the contribution due to the non-coupled level
is never exactly resonant. However, if $\left|\Delta\right|\gg\Omega$,
the light-shift, $\delta_{r}\simeq\Omega^{2}/4\left|\Delta\right|$,
can be arbitrarily small, leading to an almost resonant behaviour
if $\delta_{r}\lesssim\Gamma$. In the case where $\left|\Delta\right|\gg\Omega\gg\Gamma$
and $\Gamma\gtrsim\Omega^{2}/4\left|\Delta\right|$, we take into
account the quasi resonant contribution of the third level:\begin{equation}
\sigma_{R}=\frac{\sigma_{0}\,\Omega^{2}}{16\,\Delta^{2}}\,\left\{ 1+\frac{16\,\Delta^{2}\,\Gamma^{2}}{\Omega^{4}+16\,\Delta^{2}\,\Gamma^{2}}\right\} \label{eq:quasiresonant}\end{equation}

\end{document}